\documentclass[conference]{IEEEtran}

\usepackage{cite}
\usepackage{amsmath,amssymb,amsfonts}
\usepackage{algorithmic}
\usepackage{graphicx}
\usepackage{textcomp}
\usepackage{xcolor}
\def\BibTeX{{\rm B\kern-.05em{\sc i\kern-.025em b}\kern-.08em
    T\kern-.1667em\lower.7ex\hbox{E}\kern-.125emX}}
\begin{document}

\title{A Position- and Energy-Aware Routing Strategy for Subterranean LoRa Mesh Networks
}

\author{
\IEEEauthorblockN{
Nalith Udugampola\IEEEauthorrefmark{1},
Xiaoyu Ai\IEEEauthorrefmark{1},
Binghao Li\IEEEauthorrefmark{1},
Henry Gong\IEEEauthorrefmark{2},
and Aruna Seneviratne\IEEEauthorrefmark{1}
}
\IEEEauthorblockA{\IEEEauthorrefmark{1}University of New South Wales, Australia, \{n.udugampola, x.ai, binghao.li, a.seneviratne\}@unsw.edu.au}
\IEEEauthorblockA{\IEEEauthorrefmark{2}Roobuck Pty Ltd, Australia, henryg@roobuck.com.au}
}

\maketitle

\begin{abstract}

Although LoRa is predominantly employed with the single-hop LoRaWAN protocol, recent advancements have extended its application to multi-hop mesh topologies. 
Designing efficient routing for LoRa mesh networks remains challenging due to LoRa's low data rate and ALOHA-based MAC. 
Prior work often adapts conventional protocols for low-traffic, aboveground networks with strict duty cycle constraints or uses flooding-based methods in subterranean environments.
However, these approaches inefficiently utilize the limited available network bandwidth in these low-data-rate networks due to excessive control overhead, acknowledgments, and redundant retransmissions.
In this paper, we introduce a novel position- and energy-aware routing strategy tailored for subterranean LoRa mesh networks aimed at enhancing maximum throughput and power efficiency while also maintaining high packet delivery ratios.
Our mechanism begins with a lightweight position learning phase, during which LoRa repeaters ascertain their relative positions and gather routing information.
Afterwards, the network becomes fully operational with adaptive routing, leveraging standby LoRa repeaters for recovery from packet collisions and losses, and energy-aware route switching to balance battery depletion across repeaters.
The simulation results on a representative subterranean network demonstrate a 185\% increase in maximum throughput and a 75\% reduction in energy consumption compared to a previously optimized flooding-based approach for high traffic.

\end{abstract}


\section{Introduction}

Low Power Wide Area Networks (LPWANs), known for their low power consumption, long-range capabilities, and cost-effective deployment, are widely used in Internet of Things (IoT) applications \cite{sundaram2019survey}. 
LoRa (Long Range) is one of the most popular LPWAN technologies, typically used in LoRaWAN, a well-established, standardized protocol that enables low-power IoT devices to communicate with internet-connected gateways (GWs) in a single-hop star-type network topology.
Although LoRaWAN can be scaled across large geographical areas by deploying multiple gateways that form interconnected star clusters, this architecture relies heavily on the availability of reliable internet connectivity at each gateway.
However, in rural and subterranean environments, deploying and maintaining such LoRaWAN networks is often impractical due to limited telecommunication infrastructure and challenging terrain.
As a result, multi-hop and mesh LoRa networks, enabled by LoRa repeaters (RPs), have been emerging as a promising alternative to extend coverage in these challenging environments \cite{wong2024multihopSurvey}.
In particular, linear LoRa mesh networks, formed by arranging LoRa repeaters in a linear or sequential layout, have gained significant attention due to their suitability for large underground environments with limited internet infrastructure, such as underground mines, pipelines, and tunnels, where deployment of a LoRaWAN network with multiple gateways is not feasible \cite{branch2023linearLoRaReview, branch2020locationData, branch2020detonation, abrardo2019medievalAqueducts, wong2024multihopSurvey, cotrim2020MeshSurvey}.

In our research, we focus on subterranean networks that span complex systems of interconnected underground mine tunnels.
Despite the growing interest in underground LoRa mesh networks, many existing deployments remain custom-built for specific applications, typically using only a few repeaters and simple algorithms, without any standardized routing protocol to support dense networks with a large number of repeaters \cite{branch2023linearLoRaReview, branch2020locationData, branch2020detonation, abrardo2019medievalAqueducts, wong2024multihopSurvey, cotrim2020MeshSurvey}.

Developing a robust routing protocol for LoRa mesh networks is challenging due to many factors.
One factor is LoRa’s reliance on an ALOHA-based Medium Access Control (MAC) layer, which complicates the design and implementation of such a protocol.
Another factor is LoRa’s low data rate, which makes the use of methods that rely heavily on acknowledgments (ACKs) and control signals unsuitable.
These challenges are further exacerbated by regulatory restrictions, as LoRa devices operating in unlicensed Industrial, Scientific, and Medical (ISM) bands are often subject to a 1\% duty cycle, limiting transmissions to just 1\% of the time.
However, since our research focuses on networks operating within enclosed subterranean environments with minimal risk of interference to other spectrum users, the full 100\% duty cycle can be utilized, allowing continuous communication and maximizing network bandwidth utilization to support applications that require higher throughput.

Prior work has largely focused on adapting conventional protocols, such as Ad hoc On-Demand Distance Vector Routing (AODV)\cite{medeiros2020conventionalRoutingAnalysis, huang2018smartGridRouting, lundell2018routingProtocol}, as they are simple to implement and sufficient for very low-traffic networks operating under strict duty cycle limitations.
However, these proposals are not suitable for large-scale networks with applications that have high data demands, due to their limited scalability and high control overhead, which restricts the maximum achievable throughput.
Past work on underground linear LoRa mesh networks employs simple packet broadcasting and flooding techniques, where each repeater retransmits received messages, incurring minimal control overhead, making them better suited to support applications that demand higher network bandwidth \cite{branch2023linearLoRaReview,branch2020locationData,branch2020detonation, abrardo2019medievalAqueducts}.
Although this approach simplifies implementation, enhances network robustness, and provides better network bandwidth utilization compared to adaptations of conventional routing protocols, the retransmission of every received packet by each repeater still results in excessive network bandwidth usage and, most importantly, high power consumption, ultimately degrading overall network performance.

In our previous work \cite{ourScalabilityAnalysis}, we presented a comprehensive analysis of the scalability of underground linear LoRa mesh networks based on packet broadcasting and flooding, and proposed optimizations to support a large number of repeaters, end devices, and high traffic loads. 
In this paper, we build upon those optimisations by introducing a more intelligent routing strategy that replaces the original broadcasting and flooding mechanism. 
We focus on networks where LoRa gateways and repeaters are installed at fixed locations, while end devices may be static or mobile and located anywhere within the coverage area.

We make the following contributions:
\begin{itemize}
\item Introduce a lightweight mechanism for learning the position of repeaters that enables the estimation of their positions in underground LoRa mesh networks. 
\item Introduce a routing mechanism that uses the learned positions and enhances maximum throughput and power efficiency while maintaining a high packet delivery ratio (PDR). Specifically, our routing mechanism:
\begin{itemize}
    \item Features a standby repeater mechanism, where standby repeaters passively monitor transmissions and intervene only to recover lost or collided packets, thereby maintaining a high network PDR and low power consumption.
    \item Incorporate energy-aware routing that dynamically adjusts transmission responsibilities among repeaters based on their remaining battery levels, ensuring balanced energy consumption and effectively extending the operational lifetime of the network.
  \end{itemize}

\end{itemize}

The remainder of the paper is structured as follows: Section 2 reviews related work; Section 3 describes the proposed position learning and routing mechanisms; Section 4 evaluates performance through network simulations; and Section 5 concludes the paper.
\section{Related Work}


Cotrim and Kleinschmidt \cite{cotrim2020MeshSurvey} and Wong \textit{et al.} \cite{wong2024multihopSurvey} have surveyed recent advancements in multi-hop and mesh LoRa networks, highlighting their growing relevance and advantages over traditional LoRaWAN architecture.
These surveys highlight how mesh networks address several limitations of standard LoRaWAN, such as restricted coverage, challenges in rural or subterranean installations where the telecommunication infrastructure is scarce \cite{marahatta2020ruralMeshSmartMetering, marahatta2021ruralMeshSmartMetering, branch2020detonation, branch2020locationData}, high energy consumption under suboptimal conditions \cite{anedda2018energy, aslam2019exploring, barrachina2019towards, bezunartea2019towards}, and dependence on the main telecommunication infrastructure which limits the applicability of LoRaWAN for disaster and emergency scenarios \cite{macaraeg2020lora}.

However, there is still no standardized protocol for multi-hop LoRa networks. 
The LoRaWAN Relay standard by the LoRa Alliance is a step in that direction, but it only supports a single-hop extension between a gateway and an end device and does not support chaining multiple relays \cite{LoRaWANRelayStandard2022}.

Several studies have investigated routing protocols for LoRa mesh networks. 
In \cite{medeiros2020conventionalRoutingAnalysis}, de Farias Medeiros \textit{ et al.} evaluated conventional routing methods: Distance Vector Routing (DVR), Ad Hoc On-Demand Distance Vector (AODV), and Dynamic Source Routing (DSR), by simulating a 25-node single-gateway LoRa mesh network in an urban setting. 
Despite LoRa’s inherent limitations, such as low data rates and duty cycle restrictions, these routing protocols rely heavily on frequent control messages (e.g., HELLOs, RREQs, RREPs), route discovery processes, and acknowledgments (ACKs), all of which severely reduce available network bandwidth,  a problem that becomes particularly pronounced in larger, denser networks. 
Similarly, in \cite{huang2018smartGridRouting}, Huang et al. proposed an AODV-based routing protocol for smart grid applications, using Received Signal Strength Indicator (RSSI) values. 
However, this approach also suffers from high control overhead due to its reliance on route discovery and maintenance, making it suitable only for low data rate scenarios.

Customized Gateway-to-Gateway (G2G) routing protocols aim to extend LoRa network coverage in scenarios where some gateways lack direct Internet connectivity. 
For example, in \cite{lundell2018routingProtocol}, Lundell et al. proposed a G2G mesh routing protocol based on the Hybrid Wireless Mesh Protocol (HWMP) to extend LoRa coverage, where disconnected gateways forward packets to other gateways with internet access. 
By limiting routing responsibilities to only gateways and using custom headers, they reduced control overhead. 
Similar G2G protocols were introduced in \cite{dwijaksara2019g2gRouting} and \cite{aslam2020greenSmartCity} to improve connectivity in LoRa networks lacking full internet coverage.
However, these protocols still incur notable control overhead, lack contention management, are suitable only for a few hops, and support only uplink transmissions- limitations addressed by our proposed method.

LoRa mesh networks deployed in subterranean environments, such as tunnels and underground mines, are of particular interest to the mining and civil engineering sectors. 
However, research in this domain remains limited and typically relies on basic flooding mechanisms.
Branch et al. \cite{branch2020detonation, branch2020locationData} were pioneers in investigating linear LoRa mesh networks for underground mines, and used simple packet forwarding through LoRa repeaters where data traverses a chain of repeaters using a flooding-based approach.
Similarly, Abrardo et al. \cite{abrardo2019medievalAqueducts} implemented a chain-type linear sensor network for monitoring underground medieval aqueducts in Siena, Italy.

In our previous work on the scalability of such subterranean linear LoRa mesh networks \cite{ourScalabilityAnalysis}, we enhanced the flooding-based approach to support high data rates by optimizing repeater density, utilizing multiple frequency channels, applying faster LoRa configurations (SF7, BW 500 kHz, CR 4/5), enabling carrier sensing by repeaters, and refining random waiting periods in the algorithm.
These enhancements enable large-scale LoRa mesh networks with high traffic, where packet traversal through the repeaters remains lossless.
In this paper, we benchmark the performance of our proposed position- and energy-aware routing mechanism against our previous flooding-based approach optimized for high throughput.

\section{Approach}

This section outlines our position-learning and routing mechanisms tailored for underground LoRa mesh networks. 
The aim of the proposed approach is to improve network bandwidth utilization and power efficiency while maintaining a higher network PDR than those achievable by the conventional packet broadcasting and flooding method.

\begin{figure}[h]
\includegraphics[width=\linewidth]{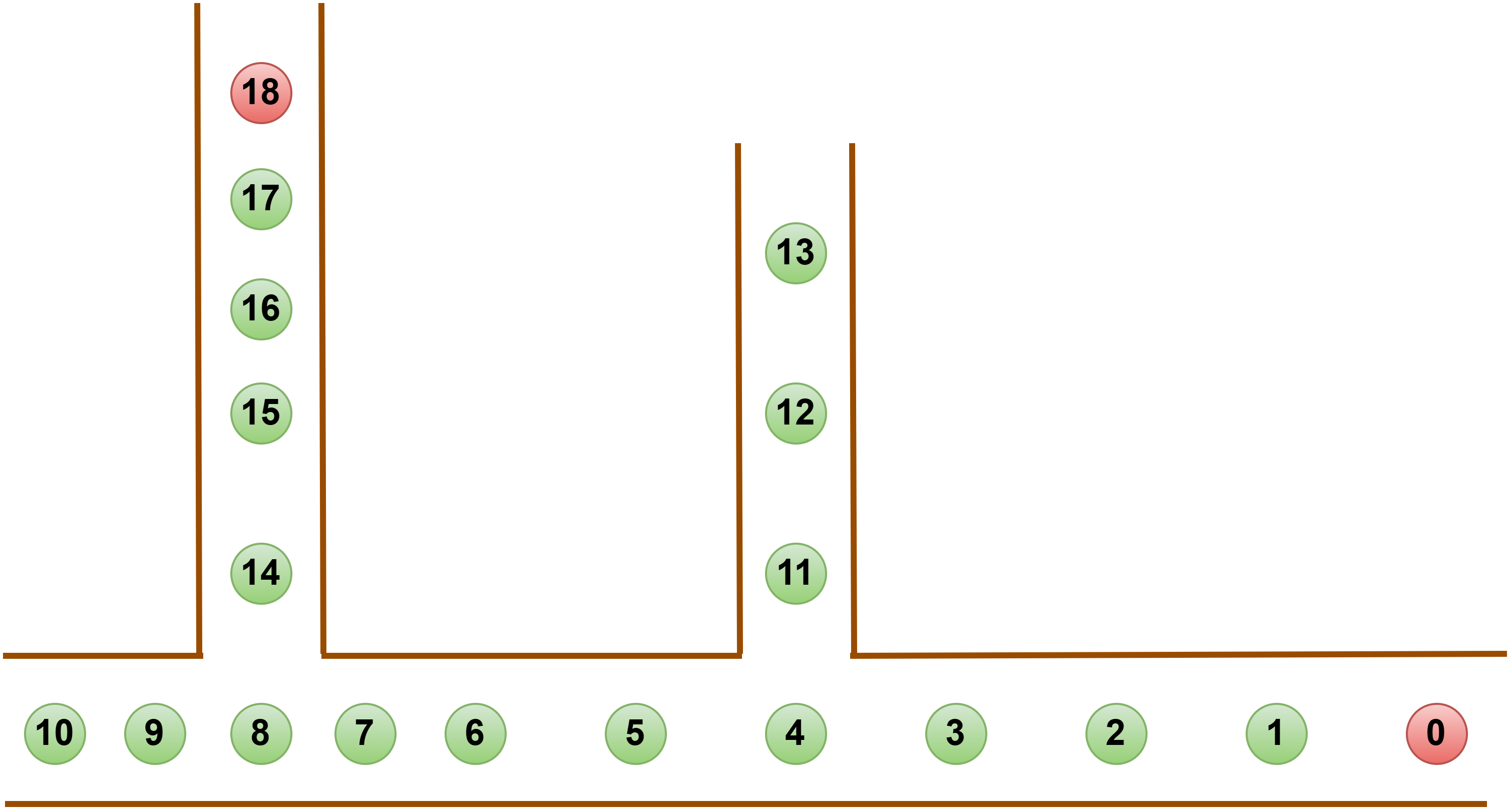}
\caption{A representative underground LoRa mesh network. Nodes 1–17 are repeaters; nodes 0 and 18 are gateways with internet access.}
\label{fig: basic representative network}
\end{figure}

To demonstrate our approach, without loss of generality, we refer to an underground LoRa mesh network shown in Figure~\ref{fig: basic representative network}. 
The numbers indicate the unique identifiers (UIDs) of the gateways (UIDs- 0 and 18) and repeaters.
Both gateways and repeaters are deployed at fixed locations.
IoT (end) devices connect to these gateways and repeaters.
For clarity, only gateways and repeaters are shown in the network illustrations; end devices are omitted.
The end devices can be static or mobile, and can be carried by people and vehicles, as well as attached to sensors and equipment.
This network includes multiple gateways and LoRa repeaters deployed along underground tunnels at varying distances that lead to different repeater densities in the tunnels.
As defined in our previous work \cite{ourScalabilityAnalysis}, repeater density is a metric that indicates the number of adjacent repeaters reachable in a single transmission in a linear LoRa mesh network. 
For example, with a repeater density of 1, a repeater transmission reaches only its immediate front and back neighbors. 
With a density of 2, the transmission reaches two repeaters in each direction.

LoRa repeaters in the network are based on LoRa concentrator modules that can simultaneously receive multiple packets with varying spreading factors, bandwidths, and frequencies. 
A repeater waits for a randomized time interval before a transmission and also performs carrier sensing with a randomized backoff to avoid packet collisions.
Also, if a repeater is in transmission mode, it will miss any incoming packets.
The network follows an asynchronous architecture with no time scheduling where end devices transmit packets on demand.
End devices transmit on a frequency channel that is separate from the one used by repeaters, preventing collisions with packets already propagating through the repeater network. 
All transmissions use the fastest LoRa configuration: Spreading factor 7, 500 kHz bandwidth, and coding rate 4/5.

Our approach consists of two main phases: a Position Learning Phase and an Operational Phase, which includes energy-aware routing and recovery logic.

\subsection{Position Learning Phase}

The primary objective of this phase is to assign a distance value to each repeater based on its proximity to the nearest gateway.

\textbf{STEP 1:} To initiate the process, all gateways broadcast special downlink packets throughout the network.
These packets flood the network, with each repeater forwarding every received packet using a common transmission power, $P_{tx}$, known to all devices.
As a result, each repeater receives transmissions from its neighboring repeaters within the communication range.
During this phase, each repeater records information about its neighbors, including their UIDs and the average received signal power, $P_{rx}$.
The received signal power serves as a coarse indicator of relative distance.

Each repeater can estimate the distance to its neighbors by assuming the well-known log-distance path loss model \cite{sharma2010logdistancemodel,branch2022goldMineModels}, given by:

\begin{equation}
L_{pl}(d) = \overline{L_{pl}}(d_0) + 10\gamma \log\left(\frac{d}{d_0}\right) + X_{\sigma}
\label{eq: log distance path loss model}
\end{equation}

 where $L_{pl}(d)$ is the path loss at a distance $d$ between the transmitter and the receiver, $\overline{L_{pl}}(d_0)$ denotes the average path loss at a reference distance $d_0$, $\gamma$ is the path loss exponent, and $X_{\sigma} \sim \mathcal{N}(0, \sigma^2)$ represents the shadowing effect, which is negligible in underground LoRa mesh networks and can therefore be ignored \cite{branch2022goldMineModels}.
 
 Since the repeater knows $P_{tx}$ and $P_{rx}$, the distance, $d$, to a nearby repeater can be calculated by substituting $L_{pl}(d) = P_{rx} - P_{tx}$ and solving equation \ref{eq: log distance path loss model} for $d$, i.e.,
 
\begin{equation}
    \begin{split}
          d 
          &= d_0 \cdot 10^{\left(\frac{L_{pl}(d) - \overline{L_{pl}}(d_0) - X_{\sigma}}{10\gamma}\right)} \\
          &= d_0 \cdot 10^{\left(\frac{P_{tx} - P_{rx} - \overline{L_{pl}}(d_0) - X_{\sigma}}{10\gamma}\right)} \\
    \end{split}
    \label{eq: distance estimation from RSSI}
\end{equation}


\begin{figure}[h]
\includegraphics[width=\linewidth]{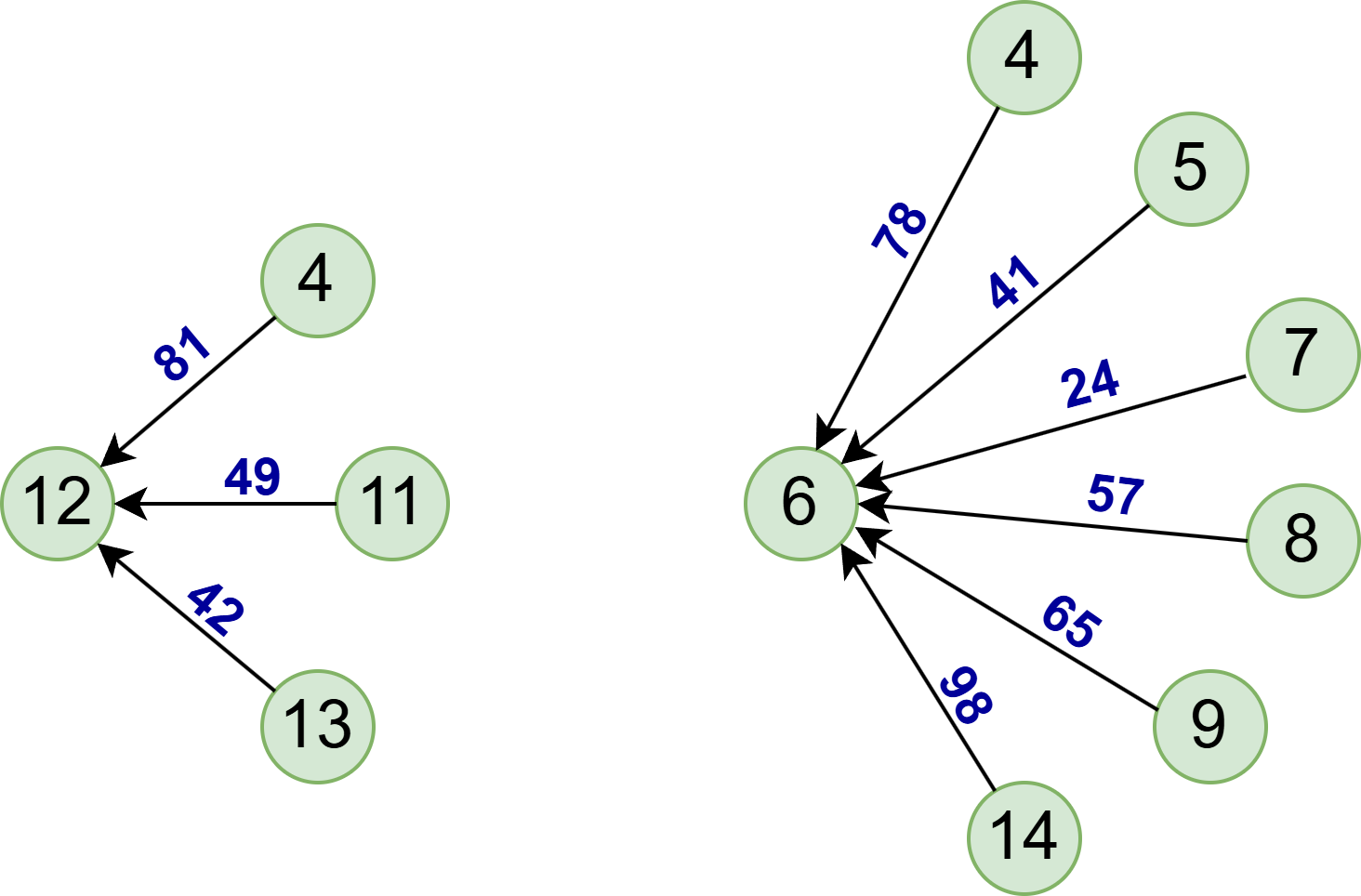}
\caption{Individual network graphs recorded by Repeater 12 and 6. The values on the arrows represent the calculated distances to the neighbours.}
\label{fig: individual network graphs}
\end{figure}

\begin{figure}[h]
\includegraphics[width=\linewidth]{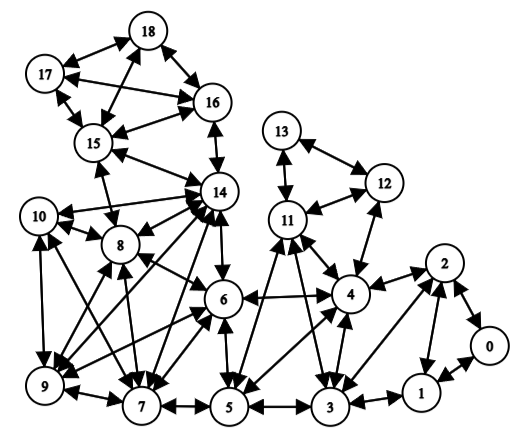}
\caption{Global network graph considering all Rx-Tx pairs}
\label{fig: global network graph}
\end{figure}

For instance, Figure~\ref{fig: individual network graphs} illustrates how repeaters 12 and 6 in our representative network (Figure~\ref{fig: basic representative network}) record the estimated relative distances to their neighboring repeaters.
These individual records by repeaters represent segments of the global network graph.

\textbf{STEP 2:} After the initial downlink broadcasts conclude, each repeater forwards its locally recorded neighborhood data to the server side by transmitting uplink packets to the gateways, using the same packet broadcasting and flooding mechanism.
Consequently, on the server side, the global network graph is then constructed by aggregating all individual neighborhood graphs, resulting in a comprehensive representation of the entire underground LoRa mesh network topology.
Figure \ref{fig: global network graph} illustrates a rough sketch of this global network graph for our example network, showing links between all transmitter (Tx)–receiver (Rx) pairs.

\begin{table}[h!]
\caption{Learned Position \& Routing Information}
\centering
\begin{tabular}{|c|c|c|c|}
\hline
\textbf{Device ID} & \textbf{Distance Value} & \textbf{Next RP/GW} & \textbf{Next RPs} \\
& & \textbf{\textit{(Upstream)}}& \textbf{\textit{(Downstream)}} \\
\hline
0  & 0   & -   & 2 \\
1  & 40  & 0   & 3 \\
2  & 85  & 0   & 4 \\
3  & 124 & 1   & 5, 11 \\
4  & 170 & 2   & 5, 12 \\
5  & 218 & 3   & - \\
6  & 199 & 14  & - \\
7  & 175 & 14  & 6, 10 \\
8  & 166 & 15  & 6, 10 \\
9  & 176 & 14  & 10 \\
10 & 194 & 14  & - \\
11 & 193 & 3   & 13 \\
12 & 242 & 4   & 13 \\
13 & 289 & 11  & - \\
14 & 123 & 16  & 6, 10 \\
15 & 82  & 18  & 8 \\
16 & 59  & 18  & 14 \\
17 & 28  & 18  & 15 \\
18 & 0   & -   & 15 \\
\hline
\end{tabular}
\label{Table: all learned info}
\end{table}

\begin{table}[h!]
\caption{Global Knowledge Learned by Repeater 3 About its Neighborhood}
\centering
\begin{tabular}{|c|c|c|}
\hline
\textbf{Device ID} & \textbf{Distance Value} & \textbf{Is Next Address?} \\
\hline
3 (self) & 126 & \\
1        & 52  & Upstream \\
2        & 76  & \\
4        & 148 & \\
5        & 188 & Downstream \\
11       & 200 & Downstream \\
\hline
\end{tabular}
\label{Table: RP3 learned info}
\end{table}

Once the network graph is generated, the server can apply Dijkstra’s algorithm to compute the shortest paths from each gateway to all repeaters, effectively forming minimum spanning trees rooted at the gateways.
Based on these paths, a distance value can be calculated for each repeater as an estimate of its distance to the nearest gateway.
Subsequently, based on the network graph and the calculated distance values, the server generates all the routing information required by the repeaters for the network's operation.

For each repeater, neighboring repeaters with lower distance values are considered to be closer to a gateway. 
Among these neighbors, the one with the lowest distance value is selected as the next uplink address (i.e., the UID of the next repeater to which data should be forwarded).

In the case of a downlink transmission, all gateways send the same downlink packet to spread throughout the network, assuming that the targeted end device could be located anywhere within the network, especially if it is a mobile node.
Repeaters that share the same nearest gateway are grouped together to form a subgraph.
Accordingly, the entire network graph can be divided into subgraphs, with the number of subgraphs equal to the number of gateways.
By running graph traversal algorithms, the server determines which neighboring repeaters each repeater should forward a received downlink packet to, in order to cover the whole area represented by the subgraph it belongs to.

Table \ref{Table: all learned info} summarizes the derived position and routing information for our example network.

\textbf{STEP 3:} As the next step, all gateways initiate a downlink broadcast to disseminate this information to all repeaters in the network, using the same packet broadcasting and flooding mechanism as before.
During this downlink transmission, each repeater extracts only the subset of information relevant to its operation, thereby gaining global knowledge of the network.
Specifically, each repeater records its own distance value, the distance values of its neighbors, and the next-hop addresses to forward downlink and uplink packets.
For example, the position and routing information extracted by Repeater 3 in the considered network is summarized in Table \ref{Table: RP3 learned info}.
This concludes the position learning phase. 

\subsection{Operational Phase with Routing}

After the position-learning phase, the network can operate more efficiently by employing the routing mechanism instead of packet broadcasting and flooding.
When forwarding a packet, a repeater simply addresses it to the next hop based on whether it is an uplink or downlink packet.
As noted earlier, uplink and downlink addresses are learned during the position-learning phase.
With routing information, both uplink and downlink transmissions are achieved with a minimal number of packet retransmissions by repeaters, resulting in lower channel utilization and reduced power consumption compared to flooding-based mechanisms, where each repeater retransmits.

Our routing approach also includes two additional features: a standby repeater algorithm and an energy-aware routing mechanism.

\subsection{Standby Repeater Algorithm}

\begin{figure}[h]
\includegraphics[width=\linewidth]{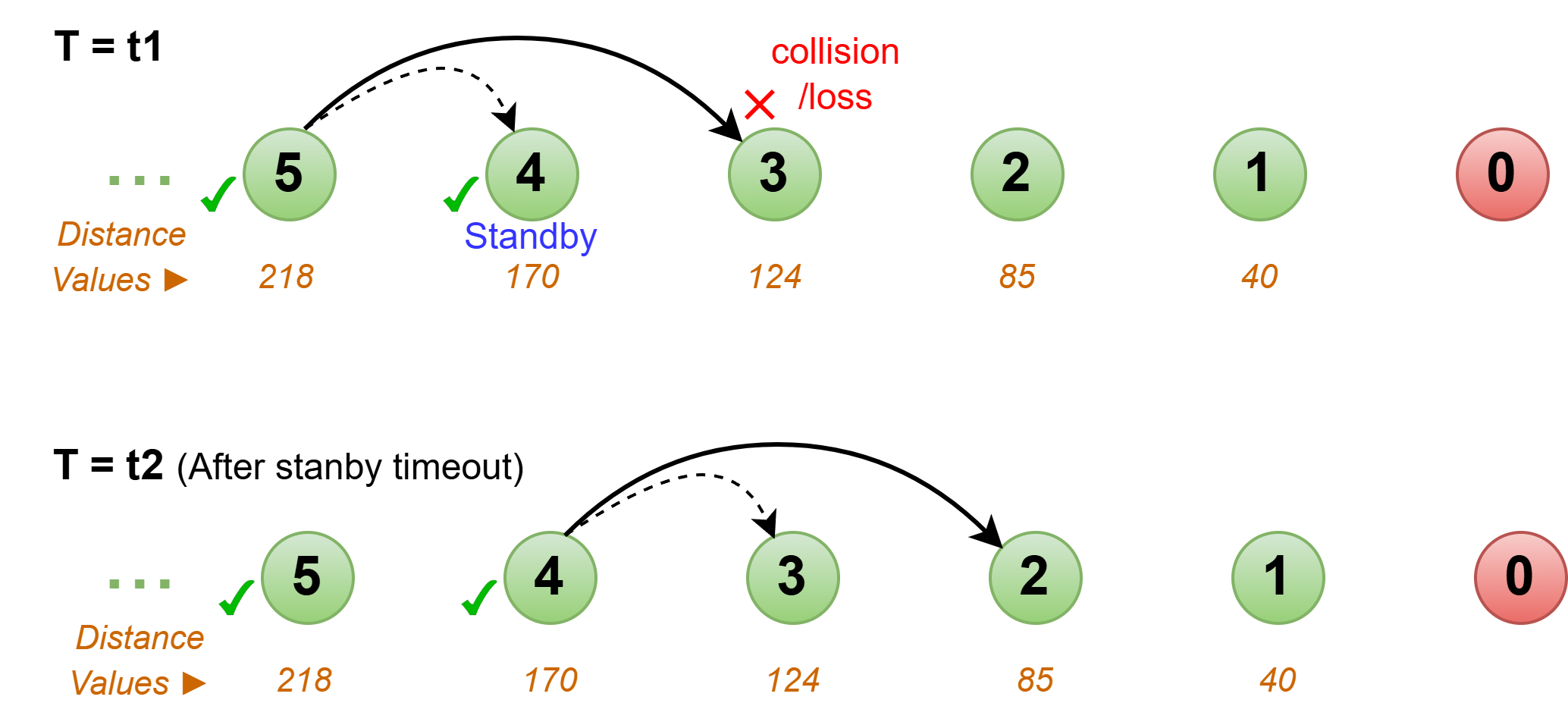}
\caption{Repeater 4 serving as a standby repeater for the uplink packet transmission from repeater 5 addressing repeater 3}
\label{fig: standby repeater functionality}
\end{figure}

As established in our previous work \cite{ourScalabilityAnalysis}, we employ a robust MAC layer where separate frequency channels are used for end devices and repeaters, with repeaters performing carrier sensing and utilizing randomized waiting periods with backoff before retransmissions to manage contention. 
As a result, any two repeaters within the same neighborhood that can hear each other will not transmit at the same time.
However, due to the hidden node problem in wireless communication, two transmissions from repeaters that cannot hear each other may still collide at a receiver in the middle. 

For example, in Figure \ref{fig: standby repeater functionality}, at time $T=t_1$, when Repeater 5 is forwarding an uplink packet to Repeater 3, neither Repeater 3 nor 4 transmits due to the carrier sensing mechanism. 
However, Repeater 5’s transmission may still collide at Repeater 3 due to simultaneous transmissions from other out-of-range repeaters, such as Repeater 1 or 2.
Despite this, due to the capture effect of LoRa \cite{rahmadhani2018CaptureEffect}, Repeater 5’s transmission is more likely to be successfully received by Repeater 4, which is closer to Repeater 5 and therefore experiences a stronger signal.

For this specific transmission, Repeater 4 acts as a standby repeater.  
When a repeater receives an uplink transmission from a neighboring repeater with a higher distance value, addressed to a neighboring repeater with a lower distance value, it operates as a standby repeater.
The standby repeater then monitors whether the intended neighbor with the lower distance value forwards the packet. 
If the packet is not forwarded, the standby repeater assumes a collision has occurred, leading to packet loss, and thus takes over the transmission to forward the packet itself.
In this example, since Repeater 4, acting as the standby repeater, does not hear Repeater 3 forwarding the packet, it assumes a packet loss. 
Therefore, after a timeout at $T=t_2$, Repeater 4 takes over the transmission and forwards the packet to its next uplink address.

In our network protocol, a repeater can function as a standby repeater for multiple transmissions simultaneously, while also receiving or forwarding packets addressed to itself. 
Furthermore, multiple standby repeaters can monitor the same packet forwarding, each with a randomly distributed standby timeout. 
In the event of packet loss, the first standby repeater to complete its timeout (i.e., the one with the shortest standby interval) will take over the transmission and forward the packet, forcing the others to back off.


\subsection{Energy-Aware Routing Mechanism}

With only the basic routing algorithm, packets generated by an end device at a fixed location would always traverse the same route.
Depending on the locations of the end devices, packet flows may become concentrated along certain routes, causing some repeaters to handle most of the forwarding, while others mostly serve as standby repeaters.
This results in imbalanced battery depletion across the network, with heavily utilized repeaters depleting more quickly, ultimately reducing the overall network longevity.
To address this issue, we incorporate an adaptive route adjustment mechanism based on the battery levels of the repeaters, aiming to achieve more balanced energy consumption and improve the network's lifespan, thereby making our routing mechanism energy-aware.

When a repeater’s battery level drops by one unit, during its next packet forwarding, it attaches its current battery level to that packet, allowing neighboring repeaters to update their records.
Thus, no additional packets are required to communicate battery levels, and this information is shared only once per battery level change for each repeater, minimizing overhead.
In our simulations, we used 100 discrete battery levels, representing the battery percentage.

\begin{figure}[h]
\includegraphics[width=\linewidth]{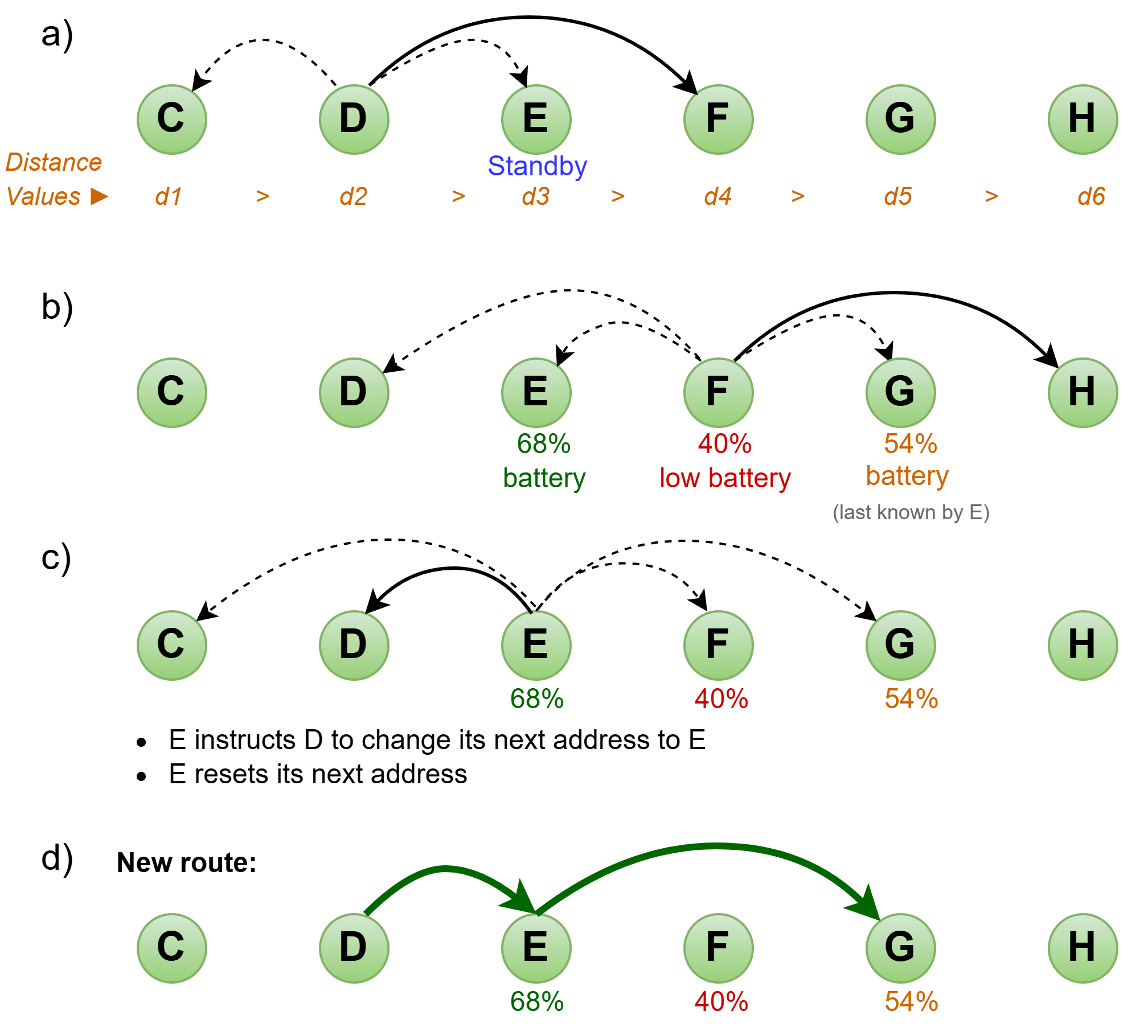}
\caption{Energy Aware Route Switching Case 1}
\label{fig: energy aware routing case 1}
\end{figure}

Route adjustments are triggered based on these battery levels under two scenarios. 
The first scenario is illustrated in Figure \ref{fig: energy aware routing case 1}, where Repeater E, acting as a standby repeater, monitors whether Repeater F forwards the packet sent by D to F.
In this case, Repeater F's battery level has recently dropped by one unit. 
Therefore, in Figure \ref{fig: energy aware routing case 1}.b, when F forwards the packet, it includes its 40\% battery level, allowing its neighbors to update their records.

While the standby repeater E confirms the packet forwarding, it also checks a specific condition: if F's updated battery level is a multiple of 10 (i.e., $\text{F's battery level}\mod10 = 0$) and both the battery levels of E and E’s original next-hop repeater (in this case, G) are more than 10 levels higher than F's battery level, E sends a special packet to D, instructing it to change its next hop address to E (Figure \ref{fig: energy aware routing case 1}.c).
If E had previously altered its next-hop address, it will also revert it back to the original address assigned during the position learning phase. 
Thereafter, as shown in Figure \ref{fig: energy aware routing case 1}.d, subsequent upstream packets forwarded by D will follow the new route, bypassing Repeater F with the low battery level, causing F to merely act as a standby repeater for these packets. 
As per the checked condition, route switching occurs only when a repeater announces a lower battery level that is also a multiple of 10, effectively limiting the frequency of energy-aware route switches and thus resulting in negligible overhead.

\begin{figure}[h]
\includegraphics[width=\linewidth]{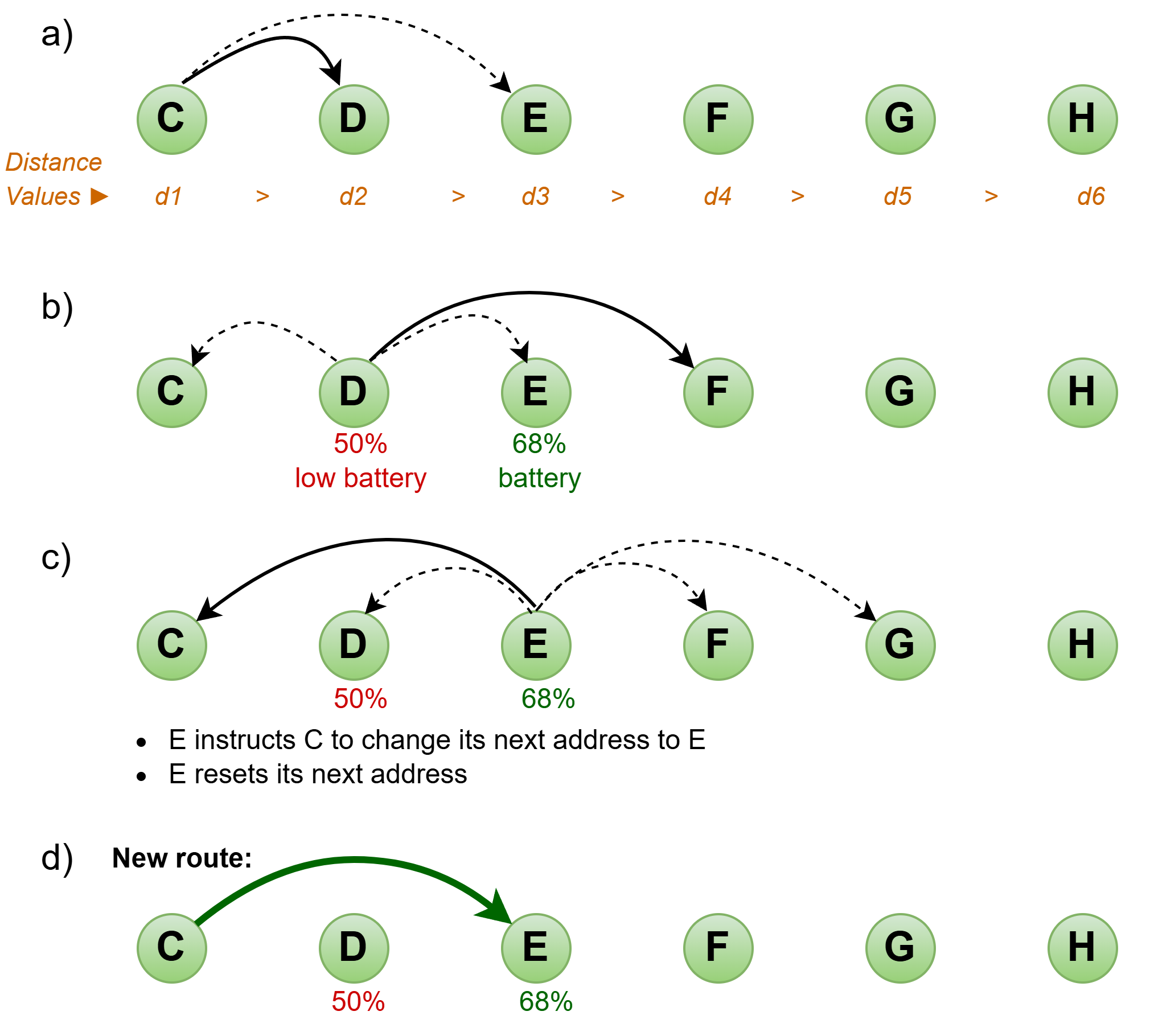}
\caption{Energy Aware Route Switching Case 2}
\label{fig: energy aware routing case 2}
\end{figure}

The second scenario of energy-aware route adjustments is illustrated in Figure \ref{fig: energy aware routing case 2}.
This scenario may arise due to prior route changes in the network. 
In this case, Repeater E overhears an upstream packet transmission from its neighbor C to neighbor D, both of which have higher distance values than E.
Additionally, Repeater D's battery level has recently decreased by one unit, so when it forwards the packet, it includes its updated battery level to notify its neighbors (Figure \ref{fig: energy aware routing case 2}.b). 
Repeater E then checks another condition: if D's updated battery level is a multiple of 10 and E's battery level is higher than D's, E sends a special packet to C, instructing it to change its next-hop address to E instead of D (Figure \ref{fig: energy aware routing case 2}.c). 
Consequently, as shown in Figure \ref{fig: energy aware routing case 2}.d, subsequent upstream packets forwarded by C will follow the new route, bypassing Repeater D with the low battery level.

Although we only describe upstream examples here, downstream transmissions follow the same logic, with the order of distance values reversed.

\section{Simulation Results}

This section presents simulation results evaluating the performance of the representative network shown in Figure~\ref{fig: basic representative network}, using both the proposed routing mechanism and our previous flooding-based approach optimized for high-throughput applications \cite{ourScalabilityAnalysis}.
To conduct these experiments, we extended our LoRaMeshSimulator framework, originally introduced in \cite{ourScalabilityAnalysis}, which consists of Python-based simulation scripts developed using SimPy, a discrete-event simulation library.
While this work focuses on the proposed routing mechanism, it builds on the core repeater logic from our previous work in \cite{ourScalabilityAnalysis}, which serves as the foundation for our network protocol.
All Python code used in the presented experiments is publicly available in \cite{ourRoutingPythonScripts}, which also serves as a complete reference for the network protocol and configuration details, supporting full reproducibility.

The experiments presented focus solely on upstream transmissions, where each end device (not shown in Figure~\ref{fig: basic representative network} for clarity) periodically transmits packets at uniformly distributed intervals with a specified mean.

\subsection{Network PDR and Latency}

In the first experiment, the end devices were evenly distributed, placing one near each repeater in the network shown in Figure \ref{fig: basic representative network}.
The network was simulated with and without the routing mechanism for 10000 upstream packets (each 20 bytes in size), with each end device periodically transmitting packets at a mean interval of 2 seconds, thereby creating a very high-throughput scenario relative to typical LoRa network conditions.

\begin{figure}[h]
\includegraphics[width=\linewidth]{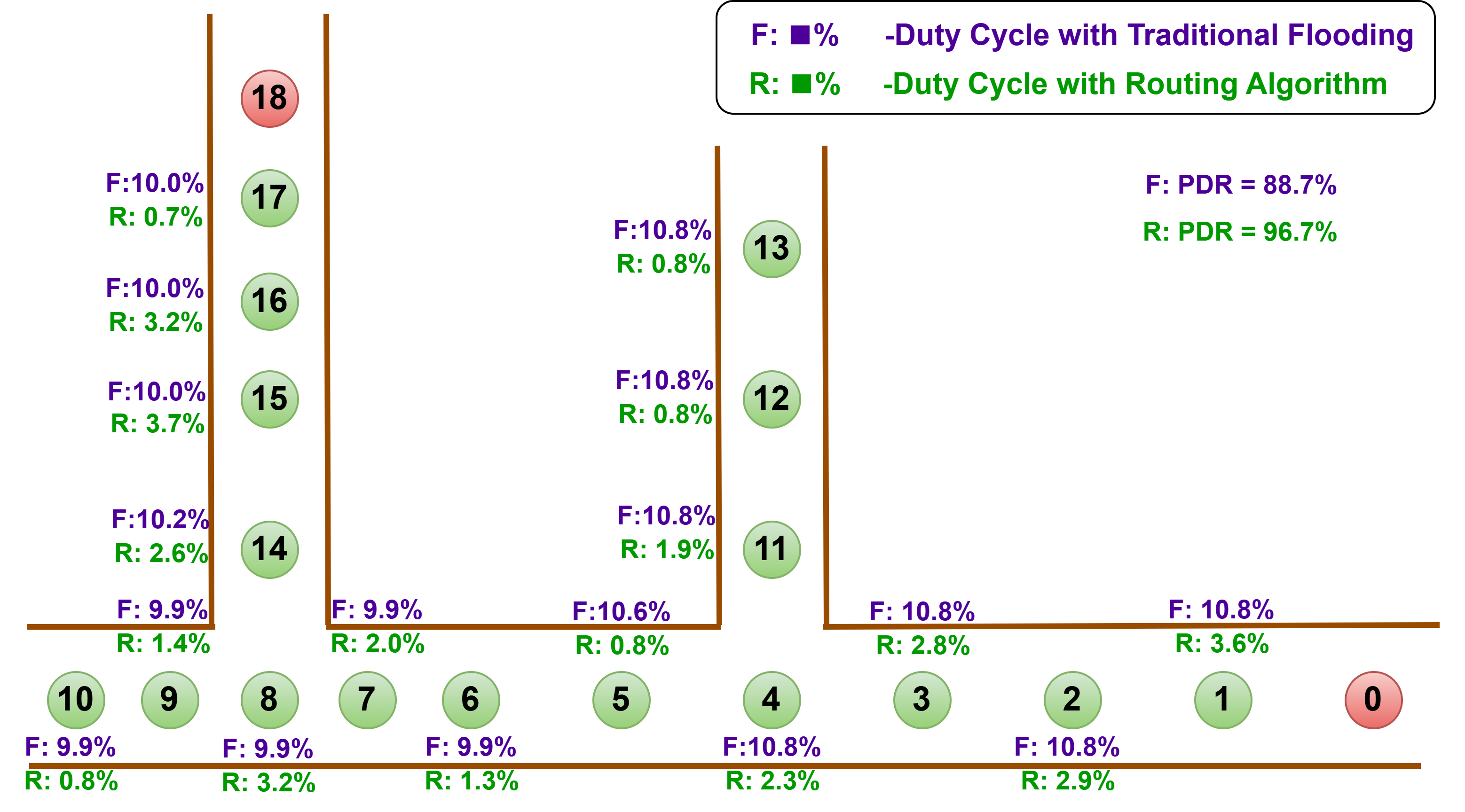}
\caption{Repeater duty cycle with and without routing algorithm}
\label{fig: repeater duty cycle}
\end{figure}

Figure \ref{fig: repeater duty cycle} illustrates the duty cycle (i.e., the percentage of time spent in the transmission state) of each repeater in the network under both the proposed routing mechanism and the flooding-based approach.
In the flooding method, since each repeater retransmits every packet duty cycles exceeded 10\% for all repeaters.
In contrast, when simulating with the routing mechanism, only the minimum number of repeaters participated in forwarding each packet, resulting in significantly lower duty cycles ranging from 0.7\% to 3.7\%.

According to Table \ref{Table: simulation summary}, which summarizes simulation results, when using the flooding-based method, the network's PDR is 88.7\%, with 1128 packets lost out of 10000.
Of these, 1061 were lost during the initial transmissions by the end devices as the repeaters they reached were in transmitting state at the time, preventing them from being received by the repeaters.
However, in the flooding-based method, when all repeaters forward the same packet, despite packet collisions, there is still a higher chance for the packet to reach a gateway, making packet losses during retransmissions (67 according to Table \ref{Table: simulation summary}) minimal.


\begin{table}[h!]
\caption{Summary of Simulation Results}
\centering
\begin{tabular}{|l|c|c|}
\hline
\textbf{Metric} & \textbf{Traditional} & \textbf{Routing} \\
& \textbf{Packet Flooding} & \textbf{Mechanism}\\
\hline
Total Generated Packets & 10,000 & 10,000 \\
Total Lost Packets & 1,128 & 329 \\
Lost at Initial ED Transmission & 1,061 & 193 \\
Lost at Intermediate Repetition & 67 & 136 \\
Overall PDR & 0.8872 & 0.9671 \\
Average Latency & 1271 ms & 581 ms \\
\hline
\end{tabular}
\label{Table: simulation summary}
\end{table}

In comparison, the simulation results for the routing mechanism, shown in Table \ref{Table: simulation summary}, indicate a network PDR of 96.7\%, with only 329 lost packets out of 10000.
Here, because of the lower duty cycles by repeaters, the likelihood of packet collisions during initial transmissions from end devices was significantly reduced.
However, with the routing mechanism, since only a few repeaters forward a packet to the nearest gateway, the number of packets lost during repetition was higher than in the flooding-based method.
Nevertheless, this increase was mitigated by the standby repeater functionality, without which the losses would have been considerably greater.
Overall, the total packet loss with the routing mechanism remains substantially lower.


Additionally, in these simulations, the average latency in the network with the flooding-based mechanism was 1271 ms, while it was only 581 ms with the routing mechanism, representing a 54\% reduction.

\subsection{Load Testing}

\begin{figure}[h]
\includegraphics[width=\linewidth]{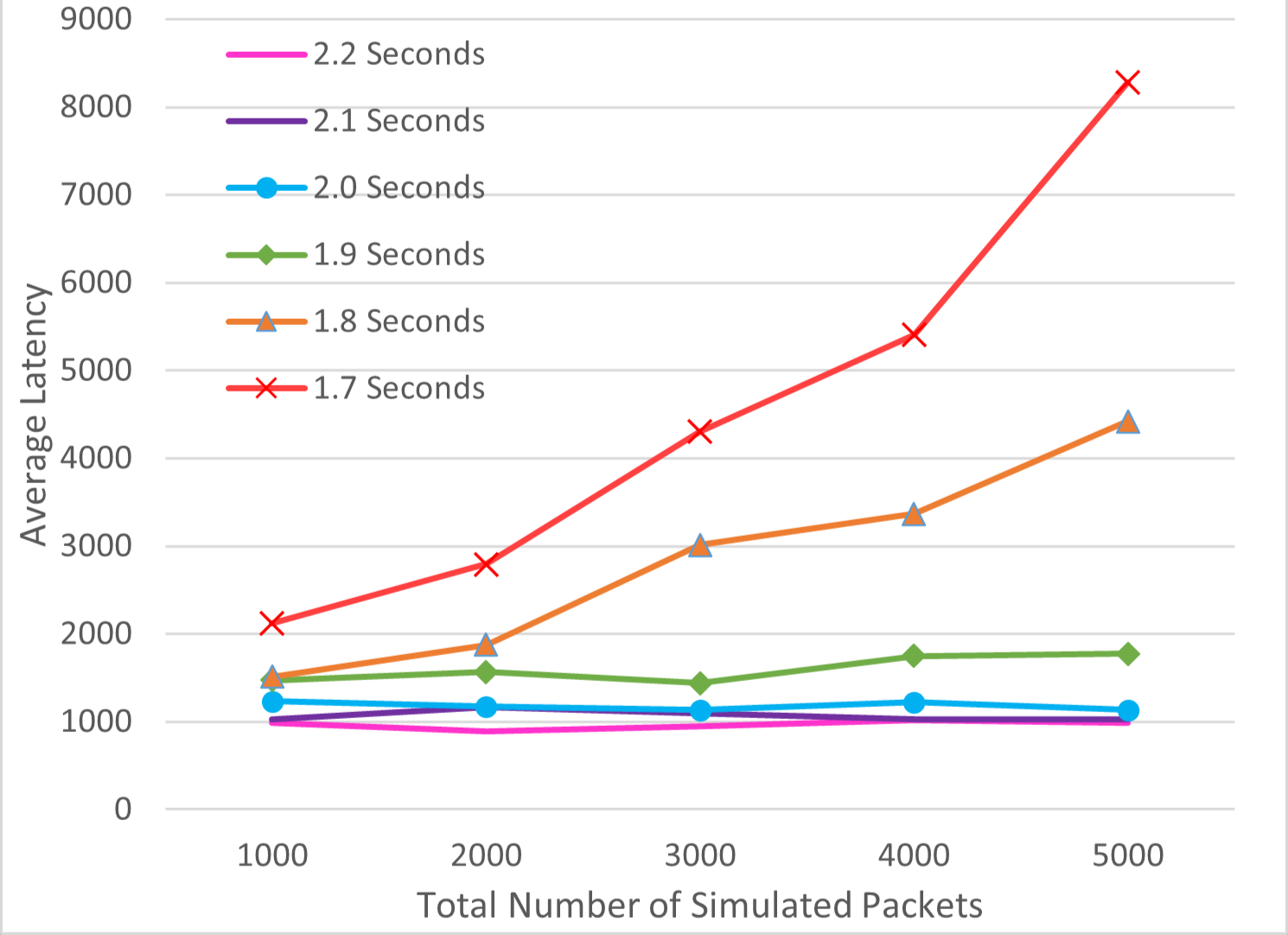}
\caption{Load testing the network with packet flooding-based mechanism}
\label{fig: load testing with flooding}
\end{figure}

\begin{figure}[h]
\includegraphics[width=\linewidth]{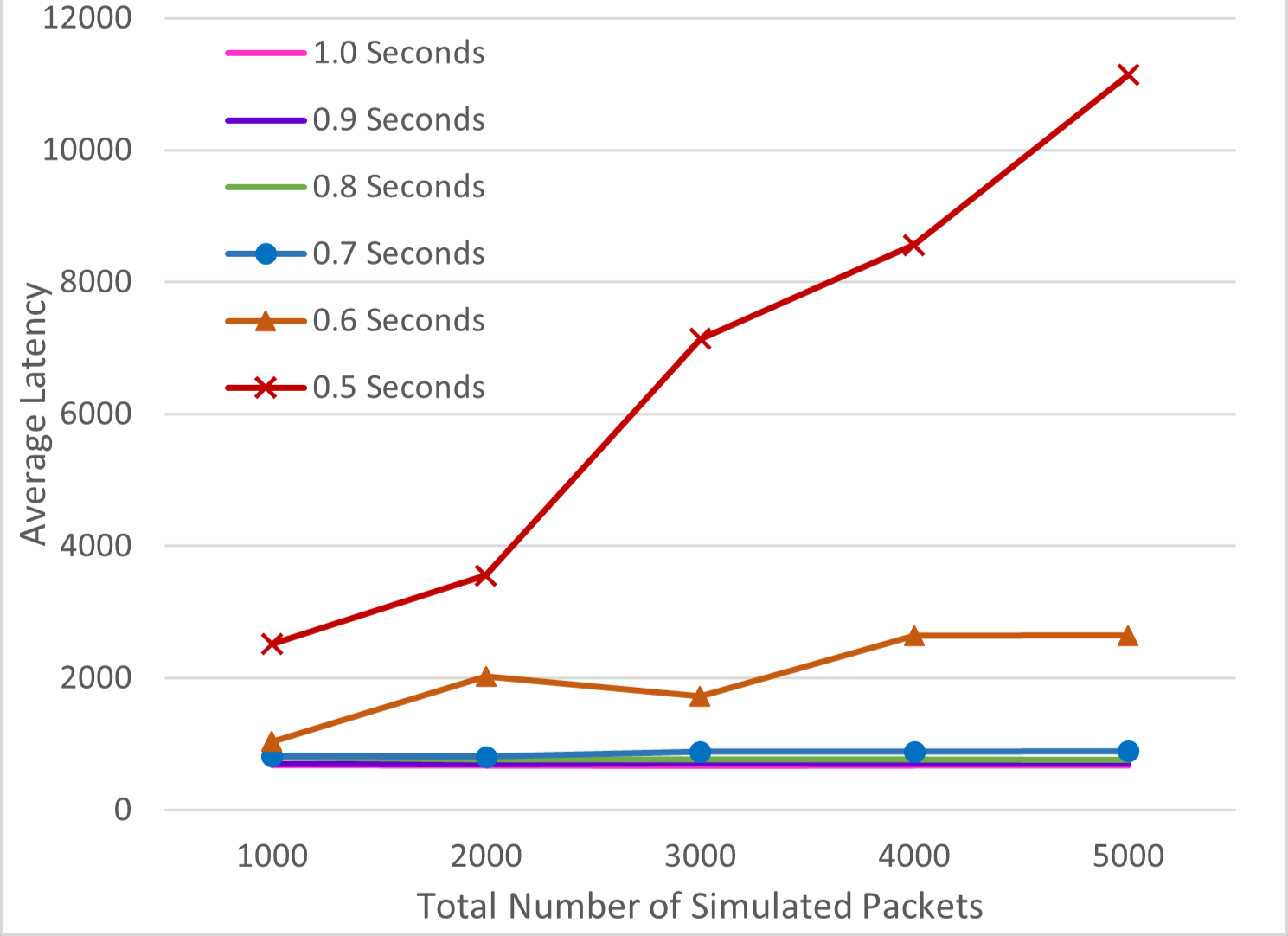}
\caption{Load testing the network with the routing mechanism}
\label{fig: load testing with routing}
\end{figure}

In this experiment, we simulated the same network arrangement while increasing the network traffic by reducing the mean time period of packet transmissions by the end devices.
As shown in Figures \ref{fig: load testing with flooding} and \ref{fig: load testing with routing}, for each packet input rate (indicated by the mean transmission period of an end device in the graphs), we conducted multiple simulations, increasing the total number of simulated packets in the network and observed the average network latency.

According to the graphs in Figure \ref{fig: load testing with flooding}, when using the flooding-based mechanism, average latency begins to increase with the number of simulated packets once the mean period of an end device transmitting packets drops below 2 seconds, instead of remaining stable.
This occurs when the Packet Input Rate (PIR) from all 17 end devices exceeds the maximum tolerable Packet Output Rate (POR) of the network.
As a result, packets increasingly queue in repeater buffers, causing network latency to rise over time.

In contrast, as shown in Figure \ref{fig: load testing with routing}, when using the routing mechanism, this rise in latency occurs only after the mean transmission interval of an end device drops below 0.7 seconds. 
Considering the PIR threshold at which latency begins to rise for both mechanisms, the proposed routing mechanism achieves a 185\% improvement in maximum throughput compared to traditional flooding in this particular network.

\subsection{Energy-Aware Routing \& Energy Consumption}

\begin{figure}[h]
\includegraphics[width=\linewidth]{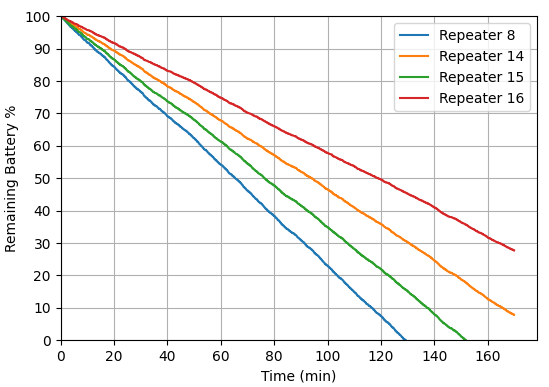}
\caption{Battery depletion over time for four selected adjacent repeaters in the network (energy-aware mechanism disabled)}
\label{fig: routing without energy aware}
\end{figure}

\begin{figure}[h]
\includegraphics[width=\linewidth]{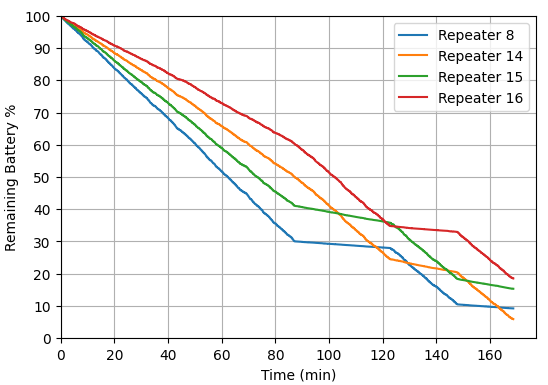}
\caption{Battery depletion over time for four selected adjacent repeaters in the network (energy-aware mechanism enabled)}
\label{fig: routing with energy aware}
\end{figure}

In this experiment, we again used the same network shown in Figure \ref{fig: basic representative network}, but placed only two end devices: one near Repeater 10 and another near Repeater 13.
Both end devices periodically transmit packets at a mean interval of 2 seconds.

For this experiment, the repeaters were assumed to be powered by 100 mAh batteries, drawing 500 mA during transmission, 50 mA while receiving packets, and operating in a low-power receive mode, consuming 1 mA the rest of the time.
This low power mode corresponds to LoRa's Channel Activity Detection (CAD) feature, as described in \cite{semtech_cad_lora}.

First, we simulated the network with a total of 10000 packets, using only the core routing mechanism, including the standby repeater function, while the energy-aware route switching feature was disabled.
As shown in Figure \ref{fig: routing without energy aware}, the battery levels of four adjacent repeaters (UIDs: 8, 14, 15, and 16) along the path to the gateway with UID 18 declined unevenly, limiting the network's operational time to approximately 130 minutes.

We then repeated the experiment with the energy-aware route switching feature enabled. 
Consequently, Figure \ref{fig: routing with energy aware} shows a more balanced battery depletion across the same repeaters, with route changes dynamically triggered based on battery levels. 
This extended the network’s operational time to over 160 minutes.

Without the energy-aware routing feature, packets consistently followed fixed routes, with some repeaters frequently forwarding packets, while others mostly remained in standby mode. 
With energy-aware route switching, repeater roles were adjusted based on battery levels, distributing traffic more evenly, and promoting network longevity with more balanced battery depletion.

Additionally, based on the current draw values mentioned above, we calculated the total power consumption of the repeaters in the first experiment with 17 evenly distributed end devices.
When using the packet broadcasting and flooding mechanism, the total power consumption was 2936 mAh, compared to just 712 mAh with the routing mechanism, resulting in a 75\% reduction in energy usage.
\section{Conclusion}

This paper presented a novel position- and energy-aware routing mechanism tailored for subterranean LoRa mesh networks, designed to support high-throughput and energy-efficient operations in challenging underground environments. The proposed strategy introduces a lightweight position-learning phase and leverages two key innovations: a standby repeater mechanism for mitigating packet losses due to collisions, and an energy-aware routing algorithm that dynamically balances energy consumption across the network.

The simulation results demonstrate that the proposed routing approach significantly outperforms an optimized packet flooding-based baseline. 
Specifically, for a representative LoRa mesh network, it achieved a 185\% improvement in maximum throughput, a 75\% reduction in total energy consumption, a 54\% decrease in average latency, and an increase in packet delivery ratio from 88.7\% to 96.7\%. 
These results underscore the effectiveness of integrating topology awareness and energy adaptivity into LoRa mesh routing design, offering a scalable and resilient communication solution for mission-critical applications in underground mines, tunnels, and other constrained environments.

\bibliographystyle{IEEEtran}
\bibliography{references}

\end{document}